\documentstyle[12pt]{article}
\def \be{\begin{equation}}
\def \ee{\end{equation}}
\def \g {\gamma}
\def \Dsl {D \!\!\!\! /}
\def \G {{\bf \Gamma}}
\def\x{{\bf x}}

\newcommand{\dd}{\raisebox{10pt}
           {\tiny$\scriptscriptstyle\longleftrightarrow$}\hspace{-11.7pt}}

\begin{document}
\setcounter{footnote}0
\begin{center}
\hfill \\
\vspace{0.3in}
\bigskip

\bigskip

\bigskip

{\Large \bf  Rarita-Schwinger Field

\bigskip
\bigskip

 in the AdS/CFT Correspondence}

\bigskip
\bigskip
\bigskip

\bigskip
\bigskip

{\Large Anastasia Volovich}

\bigskip
\bigskip
\bigskip

\begin{center}
{\it Department of Physics,\\  Harvard University,\\
Cambridge, MA 02138 \\
nastya@physics.harvard.edu \footnote{On leave from  Moscow
State University
and L. D. Landau Institute for Theoretical Physics, Moscow, Russia.}}
\end{center}

\bigskip

\end{center}
\begin{abstract}
\bigskip

A free  Rarita-Schwinger field in the Anti-de Sitter space
is considered. We show that the usual action can be supplemented
by a boundary term that  can be interpreted as the
generating functional of the correlation functions in a conformal
field theory on the boundary of the Anti-de Sitter space.
\end{abstract}

\bigskip
\bigskip
\bigskip

\newpage
\section{Introduction}
Recently it has been proposed by Maldacena \cite{M} that there is
an exact correspondence between string theory
on Anti-de Sitter space ($AdS$)
and a certain superconformal field theory (CFT) on its  boundary.
For instance, the
large N and the large t'Hooft coupling limit
of ${\cal N}=4$ four dimensional
$SU(N)$ Super Yang--Mills
can be described by  IIB
supergravity  compactified on $AdS_5 \times S^5$.
In \cite{GKP,W} the recipe for computing field theory observables via
 $AdS$ was suggested,
namely,  the action
for a field theory on $AdS$ considered as the
functional of the asymptotic value of the field on the boundary
of the $AdS$ space is interpreted as the generating functional
for the correlation functions in a conformal field theory
on the boundary. For example, for the scalar field one has
\be
\label{intro:equiv}
  Z_{AdS}[\phi_0] = \int_{\phi_0} {\mathcal{D}}\phi\, \exp(-S[\phi])
  = Z_{CFT}[\phi_0] = \left\langle \exp \left(\int_{\partial AdS_d}
  d^{d} x {\mathcal{O}} \phi_0 \right) \right\rangle,
\ee
where $\phi_0$ is the boundary value of $\phi$ and it couples to the scalar
$\mathcal{O}$ on the boundary. Hence calculating the 
action on the $AdS_{d+1}$ side allows one to obtain the correlation functions
on the $CFT_d$ side.

Recently, scalar, spinor, vector and graviton fields
have been considered on various products of $AdS$ spaces;
two, three and four-point functions were calculated
and various non-renormalization theorems suggested
\cite{GKP}-\cite{F2}.
One can deduce the scaling dimensions of operators in the conformal
field theory from the masses of particles in string theory.
Spinor fields of spin $1/2$ were discussed
in \cite{ HS} (see also \cite{MR}) where it was pointed out that the
ordinary
Dirac action vanishes on-shell; so  to get the generating functional 
in conformal field theory
one has to supplement the action by a certain boundary term,
which produces the correlation function in conformal field theory.

In this note we will discuss
the  Rarita-Schwinger field on $AdS_{d+1}$.
It can couple
to the supersymmetry current, thus
it is natural to reproduce the two-point correlation
function of the supersymmetry currents
from the partition function on the $AdS$ side by
finding the appropriate
boundary term to add to the action, since the bulk
action  vanishes onshell.
In section 2 we consider the Rarita-Schwinger field
on $AdS_{d+1}$ and find the solution of the corresponding Dirichlet
problem.
In section 3 we review  the two-point function of
operators coupling to gravitini, which is restricted by
conformal invariance.
Finally, in section 4 we find the boundary term on the $AdS_{d+1}$
side which reproduces the correlator on $CFT_d$ side, and the
relation between the scaling dimension of the
operator in the $CFT_d$  and the mass of the particle in  
the $AdS_{d+1}$ space.

\section{The Rarita-Schwinger field on $AdS_{d+1}$}

Here we will consider the Rarita-Schwinger field on $AdS_{d+1}$.
The action for the  Rarita-Schwinger
field in curved space reads
\be
\label{RSac}
I_0=\int _{AdS}d^{d+1}x\sqrt{G}\bar{\Psi}_{\mu}(\Gamma^{\mu\nu\lambda}
D_{\nu}-m\Gamma^{\mu\lambda})\Psi _{\lambda },
\ee
where
\be
\Gamma^{\mu\nu\lambda}= \Gamma^{[\mu}\Gamma^{\nu}\Gamma^{\lambda]}
,~~
\Gamma^{\mu\nu}=\Gamma^{[\mu}\Gamma^{\nu]},
\label{gamma}
\ee
$\Psi_{\mu}=(\Psi_{\mu\alpha})$ is the Rarita-Schwinger
field, $\Gamma^{\mu}=\gamma^a e^{\mu}_a,$~~
$\{\gamma^a,\gamma^b\}=2\delta^{ab}$,
 $e^{\mu}_a$ is a vielbein
such that the metric $G_{\mu \nu} = e^a_\mu e^b_\nu \delta_{a b};$
and by $[...]$ we denoted antisymmetrization of
the indices;
$\mu,\nu, a,b=0,1,...,d$.

In the coordinates $x^{\mu} = (x^0 , \x) = (x^0, x^i)$, $i = 1,
\ldots, d$, the  Euclidean $AdS_{d + 1}$ space is represented by the
Lobachevsky upper half-space
 $x^0 > 0$ and the metric $d s^2$ is given by
\be
d s^2 = G_{\mu \nu} d x^\mu d x^\nu = \frac{1}{x_0^2}
( d x^0 d x^0 + d \x \cdot d \x) .
\ee

The boundary $M_d=\partial AdS_{d+1}$ is defined by the hypersurface
$x^0 = 0$ plus a single point at $x^0 = \infty$.
With the choice of  the vielbein
\be
e_\mu^a = (x^0)^{-1} \delta_\mu^a , ~~~ a = 0, \ldots, d
\ee
for which the corresponding nonvanishing components of the
spin connection $\omega_\mu^{a b}$ has
the form \cite{HS}
\be
\omega_i^{0 j} = - \omega_i^{j 0} = (x^0)^{-1} \delta_i^j,
\ee
the operators $D_{\nu}$, $~\Dsl$ are given by
\be
D_{\nu}=\partial_{\nu}+{1 \over 2}
\omega_\nu^{b c} \Sigma_{b c}=
\partial_{\nu}+\frac{1}{2x^0}
\gamma_{0\nu},~~
\Dsl  = x^0 \gamma^0 \partial_0 + x^0
\G \cdot {\bf \nabla} - {d \over 2} \gamma^0.
\ee
Using  the condition
\be
 \Gamma^{\mu} \Psi_{\mu}=0,
\ee
we can rewrite the Rarita-Schwinger equation
\be
\label{RS-cur}
\Gamma^{\mu\nu\rho}D_{\nu} \Psi _{\rho} =
m\Gamma^{\mu\nu} \Psi _{\nu}
\ee
in the following form for $\psi _a=e_{a}^{\mu}\Psi_{\mu}$:
\be
x_{0} \gamma^{\nu} \partial_{\nu}\psi_{a} -\frac{d}{2} \gamma_0 \psi_{a}=
\gamma_{a} \psi_0 -m \psi_{a}.
\ee
The solution of the above is given by
\be
\psi_0 (x_0, \vec{x})  = \int d^d \vec{p}
e^{i \vec{p} \vec{x}} (p x_0)^{\frac{d+3}{2}}
(i \frac{\hat{p}}{p} K_{m+1/2}(p x_0)+ K_{m-1/2}(p x_0))
c_0^-(\vec{p}),
\ee
$$
\psi_i (x_0, \vec{x})=
\int d^d \vec{p} e^{i \vec{p} \vec{x}} (x_0 p)^{\frac{d+1}{2}}
(i\frac{\hat{p}}{p} K_{m+1/2}(p x_0)+ K_{m-1/2}(p x_0)) c_i^-(\vec{p})+
$$
\be
\int d^d \vec{p}
e^{i \vec{p} \vec{x}} (x_0 p)^{\frac{d+3}{2}}
[\frac{p_i \hat{p}}{p^2} K_{m+3/2} (px_0)-
\frac{ip_i}{p} K_{m+1/2}(p x_0)+
\frac{\gamma_i}{x_0 p} K_{m+1/2}(p x_0)
] c_0^-(\vec{p}).
\label{psii}
\ee
where
$p=|\vec{p}|$, $\hat{p}=\g^i p_i$; $K_{\nu}$ is the
modified Bessel function
and $c_0$,  $~c_i$ satisfy
$\gamma_0 c_{0}^-=-c_{0}^-$ and $\gamma_0 c_{i}^-=-c_{i}^-$.

The condition $\g_0 \psi_0 + \g_i \psi_i=0$
takes the form
\be
c_0^-(\vec{p})=-\frac{2ip_j c_j^-(\vec{p})}{(2m+d+1)p}, ~~~~~
\g_i c_i^-(\vec{p})=0.
\label{gauge}
\ee

Similarly, the solution of the conjugate equation is:
\be
\bar{\psi_0} (x_0, \vec{x})  = \int d^d \vec{p}
e^{i \vec{p} \vec{x}} (p x_0)^{\frac{d+3}{2}}
\bar{c}_0^+(\vec{p})
(-i \frac{\hat{p}}{p} K_{m+1/2}(p x_0)+ K_{m-1/2}(p x_0))
\ee
$$
\bar{\psi}_i (x_0, \vec{x})=
\int d^d \vec{p}
\bar{c}_i^+(\vec{p})
 e^{i \vec{p} \vec{x}} (x_0 p)^{\frac{d+1}{2}}
(-i\frac{\hat{p}}{p} K_{m+1/2}(p x_0)+ K_{m-1/2}(p x_0)) +
$$
\be
\int d^d \vec{p} \bar{c}_0^+(\vec{p})
e^{i \vec{p} \vec{x}} (x_0 p)^{\frac{d+3}{2}}
[\frac{p_i \hat{p}}{p^2} K_{m+3/2} (px_0)-
\frac{ip_i}{p} K_{m+1/2}(p x_0)-
\frac{\gamma_i}{x_0 p} K_{m+1/2}(p x_0)
] ,
\ee
with
\be
\bar{c}_0^+(\vec{p})=\frac{2ip_j \bar{c}_j^+(\vec{p})}{(2m+d+1)p},~~~~~
\bar{c}_i ^+ (\vec{p}) \g_i=0.
\ee
Let us now consider the boundary conditions on the surface
$x_0=\epsilon$, $\epsilon \to 0$ \cite{FMMR,MV2}. Without  loss
of generality we will consider the case of positive $m$ (the other case
can be treated similarly).  Then fixing the positive components
$\psi_i^{\epsilon, +}(\vec{x})$ of the
boundary field
 we find  that the negative components $\psi_i^{\epsilon, -}(\vec{x})$
vanish,
where $\psi_i^{\epsilon, \pm}(\vec{x})= \frac{1}{2}(1 \pm \g_0) \psi_i
(x_0=\epsilon, \vec{x})$. To specify solutions
we use the functions $\chi_i$
\be
\bar{\chi}_i^-(\vec{x})=
\epsilon^{m-d/2} \bar{\psi_i}^{\epsilon,-}(\vec{x}), ~~~
\chi_i^+(\vec{x})=  \epsilon^{m-d/2}
 \psi_i^{\epsilon,+}(\vec{x}).
\label{chi1}
\ee
The dependence of $c_0$ and $c_i$ on them is given by
(\ref{c_0}) and (\ref{ciii}) in appendix A.


\section{Correlator in $CFT_d$}

Conformal invariance fixes the two-point correlation function
for spin 3/2 primary fields of scaling dimension $\Delta$
up to normalization
to be \cite{cft}
\be
<O_{k \alpha}(x)O_{j \beta}(0)> =C
\frac{\hat{x}_{\alpha \beta} } {x^{2\Delta +1}}
(\delta_{kj}-2\frac{x_kx_j}{x^2}),
\label{cor_cft}
\ee
where $i,j$ are vector indices, $\alpha, \beta$ are spinor indices
and $\hat{x}_{\alpha \beta}=\g^i_{\alpha \beta} x_i.
$
\section{Boundary term}

The Rarita-Schwinger action
 vanishes on-shell; thus in order to
reproduce the two-point correlation function
of spin 3/2 fields in the boundary conformal field theory one
 has to supplement
the action by the boundary term.
This is analalous  to the spinor case \cite{HS}
and so we similarly take 
\be
\label{bt}
I \sim \lim_{\epsilon \to 0}
 \int_{M_d^{\epsilon}} d^d \vec{x} \sqrt{G_{\epsilon}}
\bar{\Psi}_{i} G^{ij}
\Psi_{j},
\ee
where as in the spinor case $M_d^{\epsilon}$
is the surface of $x_0=\epsilon$,
$G^{ij}$ is the induced metric on  $M_d^{\epsilon}$  
with determinant $G_{\epsilon}=\epsilon^{-2d}$.
To compute the total action we following \cite{FMMR}
consider the Dirichlet problem on $M_d^{\epsilon}.$

The action (\ref{bt}) can be rewritten 
in  momentum space in the form

\be
I \sim  \int d^d \vec{p} \epsilon^{-d}
(\bar{\psi}_i^+(\epsilon, \vec{p}){\psi}_i^+(\epsilon,-\vec{p})+
 \bar{\psi}_i^-(\epsilon,\vec{p}){\psi}_i^-(\epsilon,-\vec{p})).
\label{Iads}
\ee

Generally speaking the expression (\ref{Iads})
contains divergences when $\epsilon \to 0$.
Selecting the finite part, one gets the action  (see
appendix A for details of the derivation)

\be
I \sim
\int d^d p
\bar{\chi}^+_i(\vec{p}) p^{2m}
(
\frac{\delta_{ij}\hat{p}}{p}
-\frac{2(2m-1)}{d+2m-1} \cdot
\frac{p_i p_j \hat{p}}{p^{3}}
)
\chi_j^-(-\vec{p}),
\label{a_ads}
\ee
where the boundary spinors $\chi_i$ and
$\bar{\chi}_i$ are defined as in
(\ref{chi1}).

Taking the Fourier transform and integrating over $\vec{p}$ (we use
integrals from appendix B)
we get the expression for the boundary term on the $AdS$
side 

\be
I \sim
\int d^d x \int d^d y
\bar{\chi}^+_i(x)
\frac{\hat{x}-\hat{y}}{(x-y)^{2m+d+1}}\times
(\delta_{ij}-2\frac{(x-y)_i (x-y)_j}{(x-y)^2})
\chi_j^-(y),
\ee
which matches the expression for the CFT 
correlator (\ref{cor_cft}) 
with
\be
\Delta=d/2+m.
\label{delta}
\ee
So we found the agreement between the two-point function
derived from the  $AdS_{d+1}$ side and the $CFT_d$ side
once the appropriate boundary term was added.

The mass spectrum of IIB supergravity on $AdS_5\times S^5$
has been worked out in  \cite{KRN}. In particular
it was shown that  there is a massless gravitino field which
satisfies  equation (\ref{RS-cur}) with $m=3/2$.
The IIB supergravity on $AdS_5 \times S^5 $
is dual to four dimensional ${\cal N}=4$  Super Yang-Mills
\cite{M}. The supersymmetry current
$\Sigma_{\alpha}^{\mu}$
\be
 {\Sigma^{\mu}}_{\alpha A} =
       -\sigma^{k\nu} F_{k\nu}^-      {\sigma^{\mu}}_{\alpha {\dot \alpha}}
        {{\overline \lambda}^{{\dot
\alpha}}}_{A} +
 2i {\overline \varphi}_{AB} \dd \partial^{\mu}
             {\lambda_{\alpha}}^{B} + \frac{4}{3}i
 {{\sigma^{\mu \nu}}_{\alpha}}^{\beta}
             \partial_{\nu}({\overline \varphi}_{AB} {\lambda_{\beta}}^{B}),
 \ee
(see \cite{Ber}, \cite{green} for details)
which couples to the
Rarita-Schwinger field has the scaling dimension $\Delta=7/2$.
One gets this dimension from (\ref{delta}) if one takes $d=4$
and $m=3/2$. This provides a test of the AdS/CFT correspondence in this
case.
The  11d supergravity on $AdS_7 \times S^4,$
is related to the theory on M5-branes \cite{M}
and thus to (2,0) theory in the large N limit.
The spectrum of the supergravity theory was computed in \cite{PvN}
and that of  the primary operators of the
conformal algebra of arbitrary spin was discussed
in \cite{MR}, \cite{AOY}.
In particular the gravitino was discussed and the relation between
the scaling dimension and the mass agrees with our analysis.

\smallskip

After this paper was completed and prepared 
the paper \cite{Cor} appeared
with similar results.

\section*{Acknowledgements}
I would like to thank Juan Maldacena and Andy Strominger  for very
helpful discussions; and Jeremy Michelson for
useful comments.
I am grateful to Laboratoire de Physique Theorique
et Hautes Energies in Paris where part of this work was done
and especially Laurent Baulieu for invitation and Celine
Laroche for kind hospitality.
This work is partially supported by
DOE grant DE-FG02-96ER40559,
Van Vleck fellowship, Soros grant  and CNRS foundation.

\section*{Appendix A}
Here we will present the details of the imposed
boundary conditions and the derivation of the term
(\ref{a_ads}).
Let us set $x_0=\epsilon$ in (\ref{psii})
\be
\psi_i^{\epsilon}(\vec{p})= \int d^d \vec{x} e^{-i \vec{p} \vec{x}}
\psi_i(\epsilon, \vec{x})=
 (\epsilon p)^{\frac{d+1}{2}}
(i\frac{\hat{p}}{p} K_{m+1/2}(p \epsilon)+ K_{m-1/2}(p \epsilon))
c_i^-(\vec{p})+
\ee
\be
 (\epsilon p)^{\frac{d+3}{2}}
[\frac{p_i \hat{p}}{p^2} K_{m+3/2} (p\epsilon)-
\frac{ip_i}{p} K_{m+1/2}(p \epsilon)+
\frac{\gamma_i}{\epsilon p} K_{m+1/2}(p \epsilon)
] c_0^-(\vec{p}),
\ee
and introduce
\be
\psi_i^{\epsilon, \pm}=
\frac{1}{2}(1 \pm \g_0) \psi_i^{\epsilon}.
\ee
Then from the asymptotic behavior of the Bessel function
for $m>0$ we conclude that
once we fix $\psi_i^{\epsilon,+}$, then
$\psi_i^{\epsilon, -} \to 0$, so
analagous to the spinor case $\psi_i^{\epsilon,+}$
can be used to determine the boundary data and
\be
\psi_i^{\epsilon,+}(\vec{p})=
 (\epsilon p)^{\frac{d+1}{2}}
i\frac{\hat{p}}{p} K_{m+1/2}(p \epsilon)
\cdot
c_i^-(\vec{p})+
\label{e}
\ee
$$
 (\epsilon p)^{\frac{d+3}{2}}
[\frac{p_i \hat{p}}{p^2} K_{m+3/2} (p\epsilon)+
\frac{\gamma_i}{\epsilon p} K_{m+1/2}(p \epsilon)
] c_0^-(\vec{p}).
$$
Let us now find the dependence on $c_0^-$ and $c_i^-$
from $\psi_i^{\epsilon,+}$.
Suppose for some $a$ and $f$
 $$c_i^- =b_i + \frac{p_i}{p} a c_0^- + f \g_i c_0^-.$$
Then in (\ref{e}) if we require the first term in the right hand side
to be equal to the left hand side  we get
\be
b_i=-\frac{i \hat{p} \psi_i^{\epsilon,+}}{p K_{m+1/2}(\epsilon p)}
(\epsilon p)^{-\frac{d+1}{2}},
\ee
\be
a=i\frac{(p \epsilon)K_{m+3/2}(p \epsilon)}{K_{m+1/2}(p \epsilon)},
\ee
\be
f=\frac{i \hat{p}}{p}
\ee
or
\be
c_i^-=-\frac{i \hat{p}}{p}\frac{ \psi_i^{\epsilon,+}}{K_{m+1/2}(p
\epsilon)}
 (\epsilon p)^{-\frac{d+1}{2}}+
[\frac{ip_i }{p}\frac{(p \epsilon)K_{m+3/2}(p \epsilon) }
{p K_{m+1/2}(p \epsilon)}+
\frac{i \hat{p} \g_i }{p}]c_0^-.
\label{cii}
\ee

The condition (\ref{gauge}) gives the expresion for $c_0$
and $c_i$ in the form
\be
c_0^-=-\frac{2 \hat{p} p_j}{p^2}
\frac{(\epsilon p)^{-\frac{d+1}{2}}}
{(2m+d-1)K_{m+1/2}(p\epsilon)-2(\epsilon p)K_{m+3/2}(\epsilon p) }
\psi_j^{\epsilon,+},
\label{c_0}
\ee
\be
c_i^-=-\frac{i \hat{p}}{p}
\frac{(\epsilon p)^{-\frac{d+1}{2}} }{K_{m+1/2}(p \epsilon)}
\psi_i^{\epsilon,+} -
(\frac{ip_i }{p}\frac{(p \epsilon)K_{m+3/2}(p \epsilon) }
{p K_{m+1/2}(p \epsilon)}+
\frac{i \hat{p} \g_i }{p}) \times
\label{ciii}
\ee
$$ 
(\frac{2 \hat{p} p_j}{p^2}
\frac{(\epsilon p)^{-\frac{d+1}{2}}}
{(2m+d-1)K_{m+1/2}(p\epsilon)-2(\epsilon p)K_{m+3/2}(\epsilon p) }
\psi_j^{\epsilon,+}).$$

Then from (\ref{psii}) we get
\be
\psi_i (\epsilon,{\vec x})=
\int d^d \vec{p} e^{i\vec{p} \vec{x}}
[\psi_i^{\epsilon,+}+
(\epsilon p)^{\frac{d+1}{2}}
K_{m-1/2}(\epsilon p)
c_i^- -
(\epsilon p)^{\frac{d+3}{2}}(\frac{ip_i}{p})K_{m+1/2}(\epsilon p)
c_0^- ].
\label{35}
\ee
Substituting  the expression  (\ref{c_0})
and (\ref{ciii}) in  (\ref{35}) and
defining the boundary spinors
\be
\bar{\chi}_i^-= \lim_{\epsilon \to 0}
\epsilon^{m-d/2} \bar{\psi_i}^{\epsilon,-}, ~~~
\chi_i^+= \lim_{\epsilon \to 0} \epsilon^{m-d/2}
 \psi_j^{\epsilon,+},
\label{chi}
\ee
we get
\be
\psi_i =
\epsilon^{-m+d/2}\int d^d \vec{p} e^{i \vec{p} \vec{x}}
[\chi_i^{+}
-\frac{i\hat{p}}{p}\cdot
\frac{K_{m-1/2}}{K_{m+1/2}}\chi_i^{+}
-
\ee
$$
\{(\frac{ip_i}{p})\epsilon p\frac{K_{m-1/2}K_{m+3/2}-K^2_{m+1/2}}
{K_{m+1/2}}
 +(\frac{i\hat{p}\g_i}{p})K_{m-1/2}\}\times
$$
$$
\{ \frac{2p_j\hat{p}}{p^2} \cdot
\frac{1}{(2m+d-1)K_{m+1/2}-2\epsilon pK_{m+3/2}}
\chi_j^{+}\}]. $$
We assumed here that the argument of the Bessel function is $\epsilon p$.
The very similar expression can be easily written for the conjugate spinor.
Now let us proceed with calculation of the  boundary term:
\be
I  \sim
 \int d^d \vec{k} \epsilon^{-d}
(\bar{\psi}_i^+(\vec{k}){\psi}_i^+(-\vec{k})+
 \bar{\psi}_i^-(\vec{k}){\psi}_i^-(-\vec{k}))
\ee
as the functional of the boundary spinors
$\chi^+_i$ and $\bar{\chi}^-_i$:
\be
I \sim  2\int d^d \vec{k} \epsilon^{-2m}
\bar{\chi}_i^-(\vec{k})
[-\frac{i\hat{p}}{p}{\cal D}(\epsilon p) -
(\frac{2ip_ip_j\hat{p}}{p^3}){\cal R} (\epsilon p)]\chi_j^{+},
\ee
where
\be
{\cal D}(z)=
\frac{K_{m-1/2}(z)}{K_{m+1/2}(z)}, ~~~
{\cal R}(z)=
\frac{[\epsilon p\frac{K_{m-1/2}(z)K_{m+3/2}(z)-K^2_{m+1/2}(z)}
{K_{m+1/2}(z)}+2K_{m-1/2}(z)]}
{(2m+d-1)K_{m+1/2}(z)-2zK_{m+3/2}(z)}.
\ee
We notice that to get rid of singularities
one has to take terms in the square brackets
$\sim \epsilon^{2m}$.
Using the asymptotics of the Bessel
functions for small $z$:
\be
K_{\nu}(z)=z^{-\nu} 2^{\nu-1} \Gamma(\nu)-
z^{\nu} 2^{-1-\nu} \frac{\Gamma(\nu) \Gamma(1-\nu)}{\Gamma(\nu+1)}+...
\ee
we get the answer (\ref{a_ads}).

\section*{Appendix B}

We used the following integrals in the above:

\be
\int d^d p p_i e^{i \vec{p} \vec{x}} p^{\lambda} =
\frac{i A_{\lambda} (d+\lambda) x_i}{x^{d+\lambda+2}}
\ee
\be
\int d^d p p_i p_j p_k e^{i \vec{p} \vec{x}} p^{\lambda} =
\frac{iA_\lambda (d+\lambda)(d+\lambda+2)}{x^{d+\lambda+4}}
[(\delta_{ik} x_j +\delta_{ij} x_k +\delta_{jk} x_i)-
(d+\lambda +4) \frac{x_i x_j x_k}{x^2}],
\ee
where
$A_\lambda=2^{d+\lambda} \pi^{d/2}
\frac{\Gamma(\frac{d+\lambda}{2})}{\Gamma(-\frac{\lambda}{2})}$,
so that $A_{\lambda+2}=-(d+\lambda)(2+d) A_{\lambda}$.


\end{document}